\documentclass[preprint,12pt]{templates/elsarticle}

\usepackage{amssymb}

\usepackage{lineno}
\usepackage[inline]{enumitem}

\usepackage{multirow}
\usepackage{setspace}
\usepackage{titlesec}
\usepackage[tableposition=top]{caption}

\usepackage{natbib}
\setcitestyle{authoryear,open={(},close={)}}

\usepackage{amsmath}
\usepackage{gensymb}
\usepackage{color}

\journal{Icarus}

\begin{document}

\begin{frontmatter}



\title{The Origins of Asteroidal Rock Disaggregation: Interplay of Thermal Fatigue and Microstructure}

 \author[jhu,alabama]{Kavan Hazeli}
\author[jhu]{Charles El Mir}
\author[jhu,wv]{Stefanos Papanikolaou}
\author[nice]{Marco Delbo}
 \author[jhu]{KT Ramesh}

\address[jhu]{Department of Mechanical Engineering, The Johns Hopkins University, 3400 N. Charles St., Baltimore, MD 21218, USA}
\address[alabama]{Mechanical and Aerospace Engineering Department, The University of Alabama in Huntsville, 301 Sparkman Drive, Huntsville, AL 35899, USA}
\address[wv]{Mechanical and Aerospace Engineering Department, West Virginia University, WV 26506}
\address[nice]{Laboratoire Cassiop{\'e}e, Observatoire de la C{\^o}te d'Azur, B.P. 4229, 06034 Nice Cedex 4, France}

\begin{abstract}
	
  The distributions of size and chemical composition in the regolith on airless bodies provides clues to the evolution of the solar system. Recently, the regolith on asteroid (25143) Itokawa, visited by the JAXA Hayabusa spacecraft, was observed to contain millimeter to centimeter sized particles. Itokawa boulders commonly display well-rounded profiles and surface textures that appear inconsistent with mechanical fragmentation during meteorite impact; the rounded profiles have been hypothesized to arise from rolling and movement on the surface as a consequence of seismic shaking. We provide a possible explanation of these observations by exploring the primary crack propagation mechanisms during thermal fatigue of a chondrite. We present the \textit {in situ} evolution of the full-field strains on the surface as a function of temperature and microstructure, and observe and quantify the crack growth during thermal cycling. We observe that the primary fatigue crack path preferentially follows the interfaces between monominerals, leaving them intact after fragmentation. These observations are explained through a microstructure-based finite element model that is quantitatively compared with our experimental results. These results on the interactions of thermal fatigue cracking with the microstructure may ultimately allow us to distinguish between thermally induced fragments and impact products.

\end{abstract}

\begin{keyword}
thermal fatigue \sep ordinary chondrite \sep regolith \sep mechanical characterization

\end{keyword}

\end{frontmatter}


\section{Introduction}
\label{sec:introduction}

Asteroids, often referred to as minor planets, are considered to be relatively pristine and unaltered objects that preserve clues from the earliest epochs of our solar system \citep{housen1979asteroidal,michel2015asteroids,murdoch2015asteroid}. However, the exposure of asteroids to collisional evolution and the extreme space environment results in considerable surface modifications. Surface properties influence the observable traits (optical and thermal properties, physical structure, chemical and mineralogical properties \citep{sasaki2001production,brunetto2015asteroid}),  and so interpretation of the remotely collected data relies on assumptions about the key mechanisms involved. However, many of the major surface modification agents and their associated effective rates have yet to be understood. Recent missions, such as the JAXA Hayabusa mission and ESA's Rosetta mission,  provide information on and help us to understand the physical and chemical properties of the regolith covering the targeted bodies' surfaces.  \\

In September 2005, JAXA's Hayabusa mission recorded high resolution close-up images (~6 mm/pixel) from altitudes of 80 to 63 m above the surface of asteroid (25143) Itokawa, revealing a surface of blocky nature, lacking impact craters, and with the fine regolith mostly present in two areas also called smooth terrains. These are the Muses Sea and Sagamihara, coincide with low-gravitational potentials, and are generally homogeneous, featureless, and relatively flat, with particle sizes ranging from millimeters to centimeters \citep{yano2006touchdown}. In June 2011, the Hayabusa mission returned to Earth with approximately 1534 rocky particles from Itokawa \citep{krot2011bringing}. Among the returned particles, over 70\% (1087)  were monomineralic, including 580 olivine particles, 126 low calcium (Ca) pyroxenes, 56 high-Ca pyroxenes,  186 feldspars, 113 troilites, 13 chromites, 10 Ca phosphates, and 3 Fe-Ni metal \citep{nakamura2011itokawa}. X-ray microtomography was used to understand particle texture, mineralogy, and surface features in comparison with those seen in lunar soil. It was found that none of the particles contain agglutinates caused by meteorite impact, such as those seen in lunar soils \citep{tsuchiyama2011three}. Furthermore, noble gas content analysis of the grains showed that they were saturated with solar wind atoms, meaning that Itokawa's regolith is rejuvenated and has been exposed to solar wind for less than 8 million years \citep{busemann2015}.  Consequently, a surface rejuvenation process that can maintain these small fragments, ought to be active on Itokawa's surface. Further, laboratory-impact-generated fragments tend to be much more angular or faceted with cleavage surfaces, while the Hayabusa fragments are relatively smooth at similar lengthscales \citep{michikami2010shape}. These facts appear to be inconsistent with the commonly made assumption that the $mm$ to $cm$-sized rocky particles are predominantly formed by impact processes on the surface of airless bodies \citep{housen1982regoliths,horz1997barringer} similar to Itokawa. There are a number of such features observed on asteroidal regolith that are not consistent with the impact generation mechanism ~\citep{dombard2010boulders,robinson2001nature,pieters2000space,delbo2014thermal}. Here we explore the role of thermal fatigue on chondrites as an effective mechanism for rock disaggregation, in particular for larger than $mm$ objects.  \\ 

The ability of diurnal temperature cycling to cause the mechanical breakdown of surface rocks and boulders on the Earth and on other planetary bodies has been heavily debated for more than a century. Recently, studies based on field observations, laboratory experiments and modeling have confirmed the effectiveness of such thermal weathering on Earth, Mars and on other planetary bodies \citep{eppes2015cracks,dombard2010boulders,delbo2014thermal,molaro2015grain}. The surface temperature of asteroids follows a diurnal cycle with typically dramatic temperature changes as the Sun rises or sets. Thermal stresses then arise from thermal gradients and mismatches in the coefficients of thermal expansion, and these can produce driving forces on cracks in the regolith material, leading to the opening and extension of microscopic cracks. Growing cracks can lead to rock breakdown when the number of temperature cycles becomes sufficiently large. This process is known as thermal fatigue, and progressive thermal fatigue leads to thermal fragmentation \citep{delbo2014thermal}. \\

A related case of interest emerged when the European Space Agency's Rosetta spacecraft entered a close orbit about the Jupiter family comet 67P/Churyumov-Gerasimenko on 6 August 2014. The Rosetta lander, Philae, was equipped with an optical spectroscopic and infrared remote imaging system (OSIRIS) and acquired images of the surface at scales of $<$0.8 meter per pixel \citep{thomas2015morphological,bibring201567p}. The OSIRIS observations suggested that the surface features could be grouped into five categories: dust-covered terrains, brittle materials with pits and circular structures, large scale depressions, smooth terrains, and exposed consolidated surfaces. While it is stated that ``the surface of comet 67P is almost devoid of recognizable impact craters," several surprisingly large cracks  have been observed, including the 500 m long crack in the Anuket region and the 200 m long fracture in the Aker region \citep{thomas2015morphological}. The lack of concrete evidence of shear displacement along the cracks and impact sites that might produce the cracks raises the question of the mechanisms that govern surface evolution \citep{el2015fractures,thomas2015morphological}, especially given the large temperature variation and and temporal gradients likely to be experienced by the cometary surface over diurnal and  orbital time scales \citep{thomas2015morphological,delbo2014thermal}. \\

In this paper we examine the viability of the thermal fatigue mechanism for asteroidal rock disaggregation. We conduct experiments that capture \textit {in situ} the evolution of the full-field strains on the surface of a chondrite as a function of the cycling temperature and microstructure, and observe and quantify the crack growth \textit {during} thermal cycling. These observations are explained through a microstructure-based finite element model that is quantitatively compared with our experimental results.

\section{Asteroidal surface evolution and regolith formation}
\label{sec: Asteroid's Surface Evolution}

The collisional generation of superficial regoliths is a common process that modifies pristine asteroidal material. Further surface modification is also attributed to  space weathering processes from solar wind sputtering to micrometeorite bombardment \citep{pieters2000space,noble2010evidence}. Impacts evolve the surface by creating craters and ejecting the excavated material. Impact not only influences the size distribution of surface materials, but also results in spectral alteration by impact melt formation products (agglutinates) \citep{clark1992meteorite}. Other space weathering processes include irradiation, implantation, and sputtering from solar wind particles that can change physical properties such as optical spectra \citep{sasaki2001production} or microstructural features such as rock porosity \citep{tuǧrul2004effect}. Finally, thermal fragmentation due to thermal fatigue has been recently suggested as a contributing, and sometimes perhaps dominant, mechanism for the origin of regolith on small asteroids \citep{dombard2010boulders,robinson2001nature,delbo2014thermal,thomas2015morphological}. The relative contributions of these processes are important in determining the likely nature of the regolith on the asteroids. Simulations of these processes suggest at least two very different regolith formation timescales: a thermal fatigue timescale of 10\textsuperscript{4}-10\textsuperscript{6} years under specific temperature excursion rates \citep{delbo2014thermal}, and a micro-meteorite impact rate of 10\textsuperscript{8}-10\textsuperscript{9} years for a given meteorite flux and impact probability  \citep{hazeli2015regolith,el2016thermal}. These timescales are particularly important for determining the age and projected life expectancy of the surfaces of asteroids, and a thorough understanding of regolith, in terms of size-frequency distribution, shape distribution, and mechanical properties may explain the observed asteroidal surface features.\\

Rocks on the surface of asteroids experience a thermal cycle, typically with a period of a few hours. The consequences of such a temperature change are two-fold: First, any sufficiently large rock will sustain a temperature gradient, resulting in an internal stress accumulation. Second, any rock heterogeneity will develop local stress concentrations as a result of the mismatch in the coefficients of thermal expansion (CTE) for the rock's individual phases \citep{hazeli2015regolith,el2016thermal}. Hence, estimating the magnitude of the resultant stress heterogeneities requires knowledge of the thermomechanical properties for each individual phase. It has been previously shown \citep{hogan2015dynamic} that these heterogeneities play an important role in determining the resultant fragment sizes during high-velocity impacts. The current work aims at exploring the role that material heterogeneities may have in a thermal fatigue context and towards the development of internal material stress concentrations.

\section{Experimental Procedure}
\label{sec:experiment}

\begin{figure}[h!]
	\centering
	\includegraphics[width=1.0\textwidth]{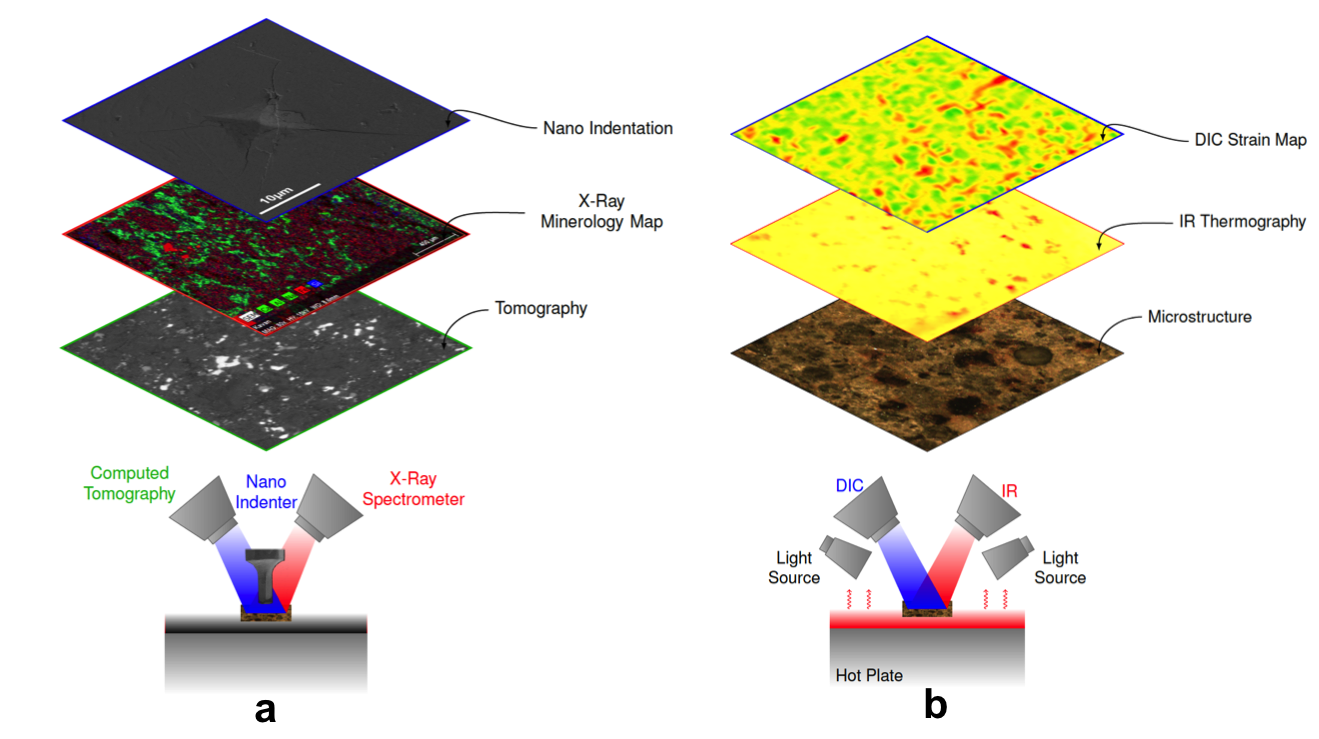}
	\caption{ Hybrid experimental set up composed of : a)\textit{ex situ} and b)\textit{in situ} measurement of the microstructure and its global and local response to the thermal cycles. (a) We used energy dispersive x-ray spectrometer to quantify major constituent phases and their surface area fractions prior to the thermal fatigue. The nanoindentation technique was implemented to measure mechanical properties of the targeted phases. (b) We integrated the Digital Image Correlation (DIC) technique with an IR camera to simultaneously record local and global strain development during thermal cycles. This approach allowed us to successfully correlate strain quantification to the corresponding phases at a given time and temperature.}
	\label{fig:experimental_procedure}
\end{figure}

 Thermal fatigue characterization was performed on an L6 ordinary chondrite (GRO 85209) that was found in the Grosvenor Mountains, Antarctica, and provided by the Smithsonian. Our experimental procedure combines state-of-art techniques for chemical, mechanical, and thermomechanical characterization of materials. Combining several characterization techniques ({\it cf. Figure~\ref{fig:experimental_procedure})} with computational modeling was necessary to understand the mechanisms associated with developing internal stresses and crack preferential path under thermal cycling. In addition to developing a reliable thermomechanical model to simulate the response of the sample used in the experiments, it was important to identify some key variables and model parameters, including the sample's microstructural components, their area fractions, and their associated mechanical and thermal properties.\\

 \subsection{Computed Tomographic measurements}
\begin{figure}[h!]
	\centering
	\includegraphics[width=1\textwidth]{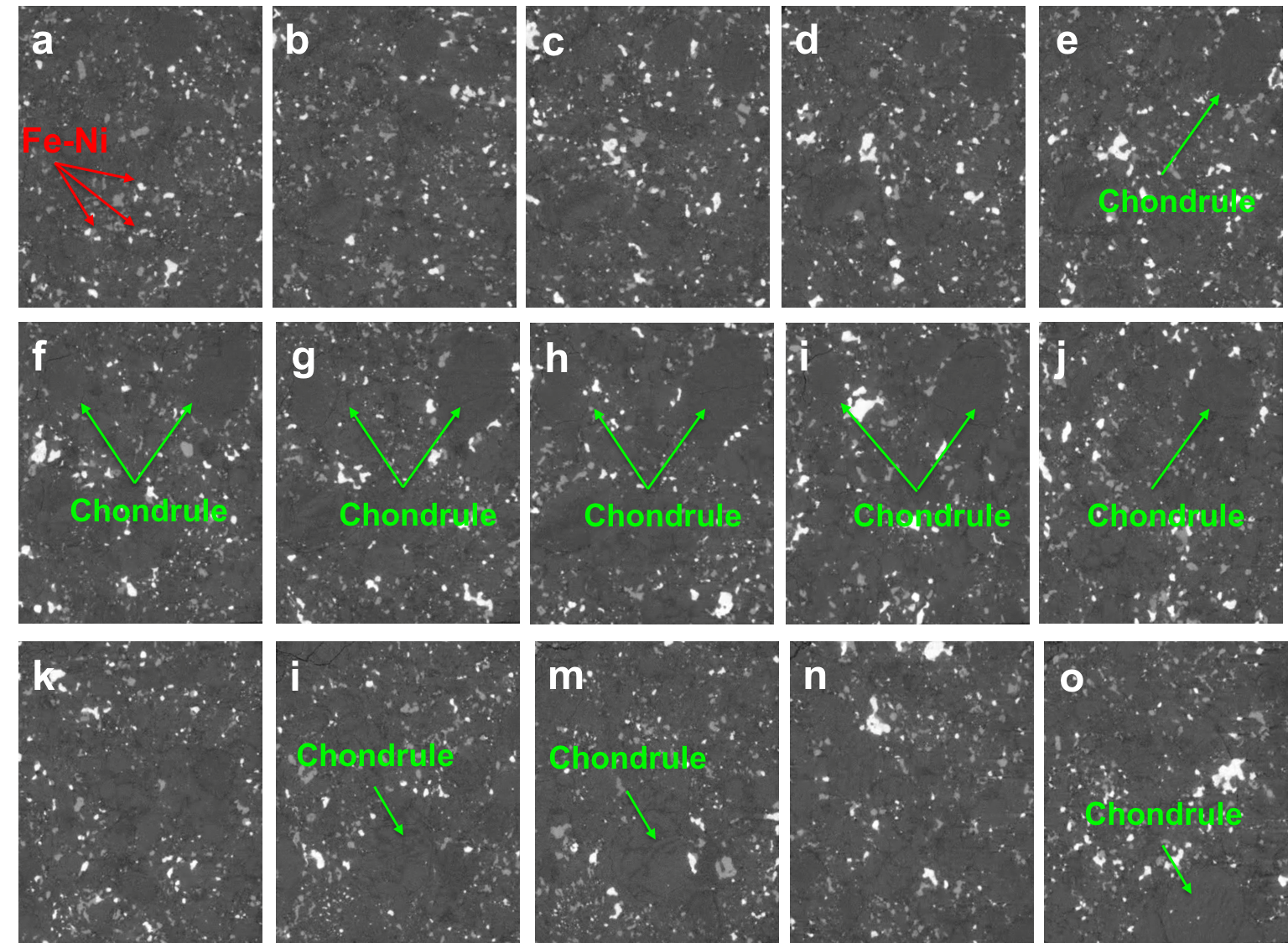}
	\caption{ X-ray CT images to study spatial structural elements distribution in a L6 chondrite. The presented serial section imaging alongside with the minerologogical composition analysis shows that the studied material (GRO 85209) is mainly composed of Fe-Ni, Chondrule, plagioclase and Mafic (last two are considered as matrix in this study).  Image (a) to (d) demonstrate Fe-Ni distribution throughout the volume. Chondrule structure starts to appear in snapshot (e) and by going through thickness (images (f) to (k)) we can observe chondrule structural evolution through the thickness of the sample. Images (i) to (m) are used to show chondrule distribution in a different region of the sample.}
	\label{fig:xray_ct}
\end{figure}

 Computed Tomographic (CT) measurements enable us to visualize the internal features within the non-transparent meteorite sample, giving insight into the volumetric size and shape distribution of the underlying microstructure. This is important because it assures what we see and measure on the surface is a good representative of the material. We analyzed 14 samples ($5 mm^3$ each) with synchrotron x-ray microtomography ($\mu$CT) using the GSECARS 13-BMD beamline at the Advanced Photon Source (APS) at the Argonne National Laboratory. We used the imaging setup described in \citep{ebel2007meteorite} using techniques similar to those employed in \citep{friedrich2008three}. During the imaging experiments, we used monochromatic x-rays of 46 keV energy and a spatial resolution of 6.12 $\mu$m/voxel edge.  Tomography measurement was used to assess and to identify the distributions and configurations of phases in the tested L6 chondrite throughout the volume. Selected CT image slices  are shown in Figure~\ref{fig:xray_ct}, starting from the top and moving towards the bottom of the specimen to visualize the spatial distribution of the microstructural features, including Fe-Ni and chondrules. These CT measurements reveal that the through-the-thickness variation in the area fraction distributions is not significant, and thus the features observed on the surface are reasonably representative of the underlying microstructure.\\

\subsection{Nano-indentation and local mechanical properties of phases}
 The CT investigation showed that the sample porosity was about 3\%. The porosity and heterogeneity distributions impose restrictions on the techniques that can be used for mechanical characterization of the sample. If one wishes to obtain overall properties for the chondrite, one must make measurements with sample volumes sufficiently large to provide representative behavior. On the other hand, the local properties associated with individual phases require a measurement technique that has sufficient spatial resolution to examine each phase. Since our interest is partly in the effects of the microstructure on the crack propagation, we need to know the properties of the individual phases, and so we used the technique of nanoindentation that allows us to examine the mechanical properties of $10\mu m$-sized inclusions with accuracy. In particular, we used the continuous stiffness measurement (CSM) technique to measure the elastic modulus and hardness of the major constituent phases in the meteorite (\textit{see Table~\ref{tab:propertiestable}}). Note that using the CSM technique makes it possible to measure the mechanical properties of the phases without the need for discrete unloading cycles, and with a time constant that is at least three orders of magnitude smaller than the time constant of more conventional methods for determining the stiffness from the slope of an unloading curve \citep{oliver2004measurement,li2000continuous}. Measurements can be recorded at exceedingly small penetration depths (tens of nanometers), making this method virtually nondestructive to the original material.

\begin{table}[ht] 
	\centering
	\caption{Mechanical Properties of the Major Phases in the Chondrite}
	\label{tab:propertiestable}
	\resizebox{\columnwidth}{!}{	
		\begin{tabular}{lccccc}
			\hline
			Material         & Bulk Modulus (GPa)     & Shear Modulus  (GPa)    & Poission's Ratio     & Coefficient of Thermal Expansion   (K\textsuperscript{-1})       & Ref.                   \\
			\hline
			Olivine             & 174               & 95                            &0.25             &$3\times10$\textsuperscript{-6}      &\cite{spray2010frictional} \\
			Pyroxyne         & 108              & 92.50                        &0.23             &$8\times10$\textsuperscript{-6}      &\cite{spray2010frictional} \\
			Plagioclase     & 83                & 8.99                          &0.33               &$4\times10$\textsuperscript{-6}      &\cite{molaro2015grain} \\
			Chondrule       & 83                & 50                        &0.25              &$10.4\times10$\textsuperscript{-6}   &This study \& \cite{delbo2014thermal}\\
			Mafic               & 136             & 8.99                           &0.24              &$6.4\times10$\textsuperscript{-6}    &This study \\
			Matrix              & 125             & 8.99                           &0.31              &$4.3\times10$\textsuperscript{-6}     &This study\\
			(Plagioclase20\%-Mafic80\%)         \\
			Fe-Ni               & 175            & 8.99                               &0.30             &$12.6\times10$\textsuperscript{-6}   &This study\\
			(Fe95\%-Ni5\%)         \\
			\\                       
\hline
\end{tabular}
}
\end{table}

\subsection{X-ray mineralogy}
\begin{figure}[h!]
	\centering
	\includegraphics[width=1\textwidth]{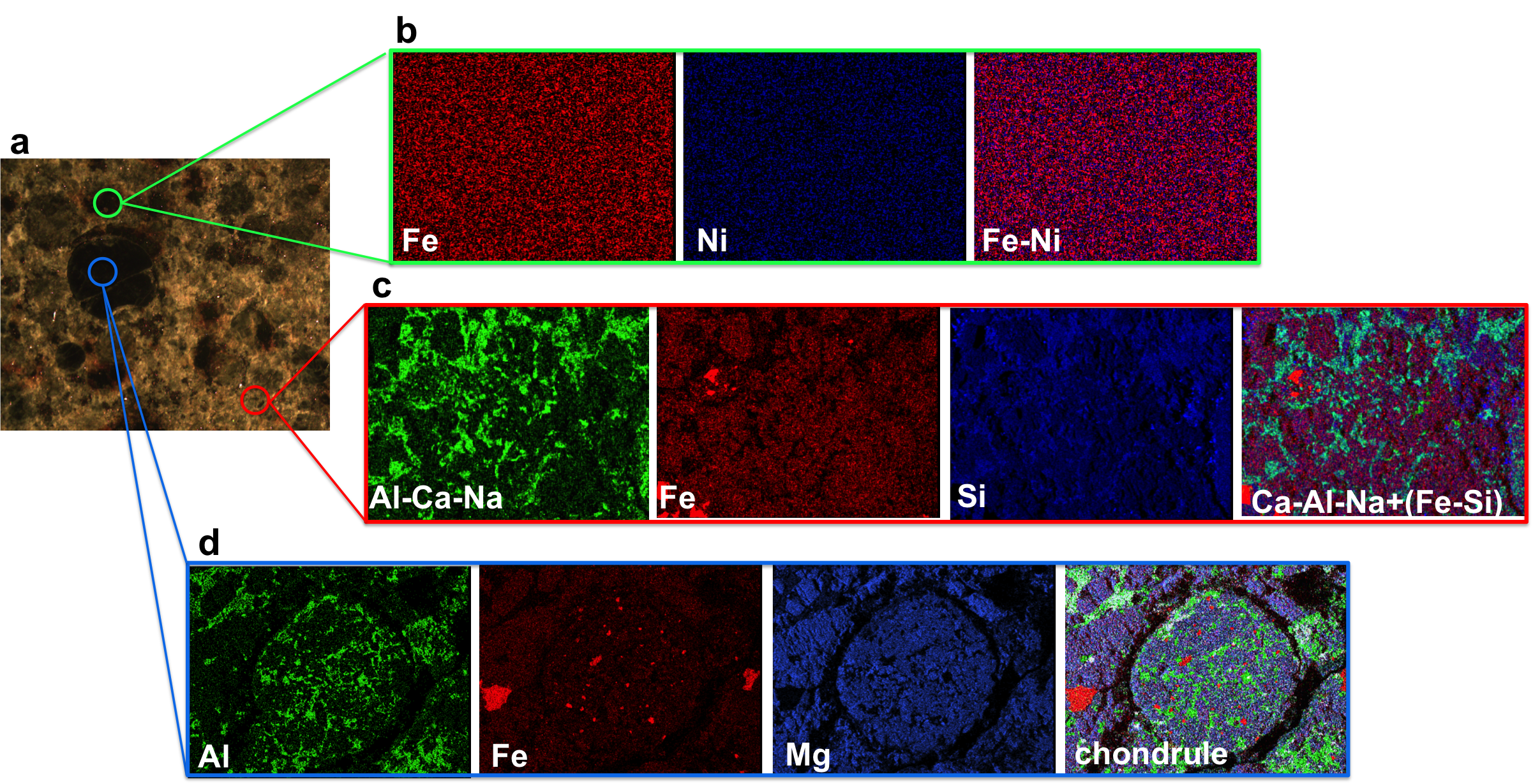}
	\caption{ Optical micrograph of the L6 chondrite (GRO 85209) supplemented with corresponding compositional mapping obtained by energy dispersive x-ray spectrometer.  Qualitative maps presented here suggest that the main large heterogeneities are Fe-Ni, and chondrules. It can be seen that the matrix is composed of Plagioclase, Olivine , and Pyroxene.}
	\label{fig:optical_micrograph}
\end{figure}

 Once the major distributions were determined using CT measurements, an energy dispersive x-ray spectrometer (EDS) was used to obtain quantitative mineralogical composition maps. Figure~\ref{fig:optical_micrograph} presents elemental and mixed data for the constituents. We used an image analysis algorithm to calculate the area fraction of each phase. It is seen that the tested L6-chondrite is mainly composed of three phases: Iron-Nickel (Fe-Ni(95\%-5\%)), chondrules, and a bulk matrix that consists of Plagioclase(20\%)+Olivine(40\%)+Pyroxene(40\%).

\subsection{Thermal Strain Measurements}

In order to understand the response of the ordinary chondrite microstructure to the thermal cycling, we measured the evolving strain fields developed \emph{during} the thermal cycling.Note that our interest in this experiment is not so much to measure the rate of crack growth as it is to determine the influence of the microstructure/mineralogy on the crack path. Thus simultaneous access to the strain field, images of the microstructure, and the temperature field is essential. As shown in Figure~\ref{fig:experimental_procedure}b, the thermal cycling experiments were coupled with digital and infrared thermography cameras to capture the strain evolution and to simultaneously record the full-field temperature map. Infrared thermography was conducted using a FLIR A325sc thermal camera with a spectral range of 7.5-12 µm, 320 x 240 pixel resolution and a maximum acquisition rate of 60 Hz allowing real-time tracking of surface temperatures.  Single 5 mega pixel camera was used to acquire images of the specimen surfaces while it was thermally cycled. The Digital Image Correlation (DIC) technique was utilized to measure the evolution of the strain fields as a consequence of both the global temperature change and the local coefficient of thermal expansion (CTE) mismatches between phases during thermal-cycling. The meteorite was subjected to thermal cycling in air using a programmable hot plate at a rate of 2\degree C/min from room temperature, 28\textsuperscript{o}C, to 200\textsuperscript{o}C and then back to room temperature. This temperature rate is considered to be representative of the thermal rates calculated for some NEA surfaces \citep{delbo2014thermal}. \\

 Measuring strain using the DIC method relies on measurements of surface deformation by comparing an original (reference) configuration with subsequent states through tracking contrast changes, which are typically achieved by applying a speckle pattern on the tested sample  (e.g. by spraying paint markers on the surface). In contrast to most DIC measurements, no artificial speckle pattern was required in our case due to the natural high-contrast microstructural features in the meteorite sample. \\
 
\begin{figure}[h!]
	\centering
	\includegraphics[width=1.0\textwidth]{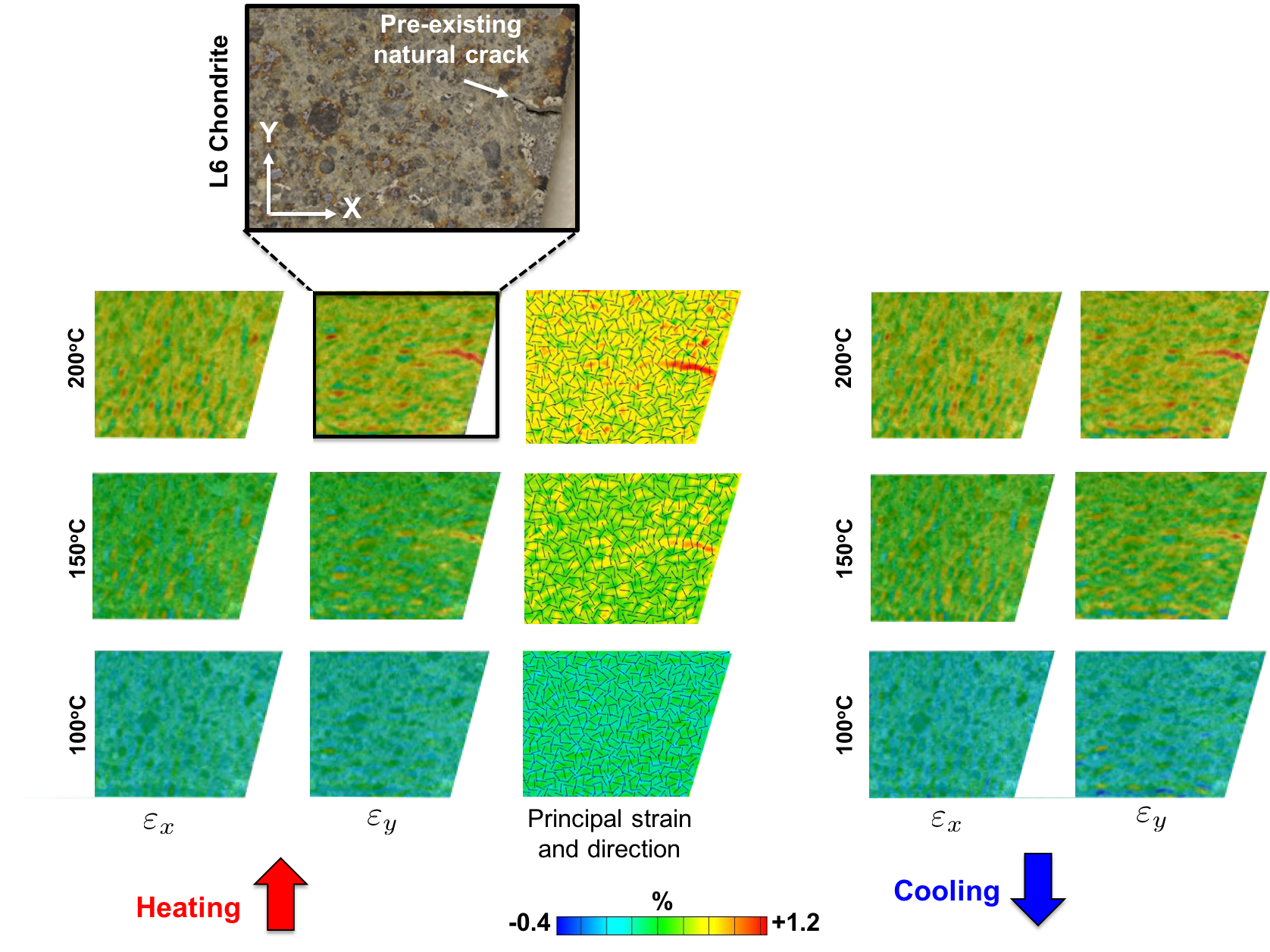}
	\caption{ L6 meteorite microstructure and its associated full-field strain map at different temperatures for one full thermal cycle (heating/cooling). The strain map develops in a heterogeneous pattern as a result of the different CTE values of each individual constituent. Average full field major principal strain computed directly from \textit{in situ} measured strain components along X and Y direction as a function of temperature is also presented.}
	\label{fig:strainmaps}
\end{figure}

The tested sample dimension was $30 \times18$ $mm$, and 6$mm$ thick. The sample's surface was polished down to $1\mu$m using an alumina suspension. A photograph of the tested specimen is shown in Figure~\ref{fig:strainmaps}: this sample had a naturally developed pre-existing crack at the middle of the right hand side of the image. The presence of this natural crack provided a unique opportunity to monitor the crack propagation path and the associated strain fields as a consequence of thermal fatigue. However, note that our optical system does not have the resolution to measure the crack-tip strain fields (which have strong gradients) themselves.

Figure~\ref{fig:strainmaps} shows the evolution of the full fields of two different components of the strain tensor, at different temperatures, during a single full cycle (heating/cooling) out of the hundred cycles performed on this sample. It is evident, when the meteorite microstructure is overlaid on the strain fields, that the measured strain fields are directly correlated with the temperature-induced local displacements of the constituents. It is also observed that the local strain evolves heterogeneously with temperature, and this heterogeneity is strongly seen at the higher temperatures. This observation is consistent with the thermal strains arising from the mismatch of the coefficients of thermal expansion (CTE) values between the various phases. The differential strain essentially scales with $\Delta \alpha \Delta T$, where $\Delta \alpha$ is the CTE mismatch and $\Delta T$ is the change in temperature. 

Finally, we use the strain components measured to compute the magnitude and direction of the principal strains during the thermal cycling (this is also shown in Figure~\ref{fig:strainmaps}). The spatial variation of the principal strains and the principal directions that is observed is a direct result of the heterogeneous character of the meteorite for this kind of simple uniform thermal cycling condition. A series of images showing the major principal strain and the corresponding principal direction (arrows) are shown in Figure~\ref{fig:strainmaps} for one heating cycle, demonstrating that the major principal strain distributions evolve with temperature (from 100\textsuperscript{o}C to 200\textsuperscript{o}). By overlaying the principal direction on top of the principal strain maps, we see that the orientation of the major principal strain (near the preexisting crack) rotates with increasing temperature to line up normal to the crack growth direction. This is what we would expect if the crack grows, since the crack tip opening displacements would generate such an apparent principal strain at the resolution of these images. The in situ thermal and DIC images thus suggest that the crack is growing as a consequence of the uniform thermal cycling of the meteorite sample.

\section{Thermomechanical model}
\label{sec:thermomech}

The extensive characterization that we performed on the different phases in our sample allows us to perform an experimentally-informed numerical simulation of the response of the material to variations in the temperature. Note that our DIC recordings allow us to measure strains, but these do not easily translate to stresses since thermal expansion could occur without resulting in any internal stresses. A numerical model that is informed by the experimental measurements is therefore required to compute internal stresses. The stresses themselves are needed to provide some indication of the reasons for the specific crack propagation path.
 
Using high-resolution images of the meteorite's surface, we constructed a microstructurally adaptive finite element mesh using an object-oriented finite element code (OOF2)\citep{langer2001oof}. Since the CT scans ({\it cf. Figure~\ref{fig:xray_ct}}) showed a nearly uniform volumetric variation indicating that the surface features are representative of the volumetric features, we believe that a 2D plane stress approximation is justified and  provides sufficient information for the purpose of this study.

A "skeleton" consisting of a regular array of large elements is first imposed on the microscopic image. The coarse skeleton mesh is then refined through a series of adaptive methods by either adjusting the mesh nodal positions or by inserting new elements, until the mesh conforms (to specified accuracy) with the assigned image of the underlying microstructure. This adaptive process is efficiently performed by defining an effective potential, which is minimized as the mesh complies with the background image, while at the same time maintaining a set of well-shaped elements that do not introduce erroneous numerical artifacts due to irregular sharp or incompatible elements \citep{reid2009modelling}. Hence, this potential is a combination of a homogeneity potential related to how many different materials, or pixels, exist within the same element, and a shape potential related to the quality of the shapes of elements. The effective potential is therefore expressed as:
 
 \begin{align} 
 U_{eff}&=\beta{U_{hom}}+(1-\beta)U_{shape}
 \end{align}
 
 \noindent where $U_{hom}$ and $U_{shape}$ represent the homogeneity and the shape potential of the mesh elements respectively, and $U_{eff}$ is the resulting effective potential. The parameter $\beta$ is a weight whose value ranges between 0 and 1, depending on whether an improved shape quality or an improved homogeneity of the mesh elements is preferred.

Once the finite element mesh was generated and aligned to conform with the actual microstructure of the sample, we assigned the material properties (shown in Table~\ref{tab:propertiestable}) that were measured in our experiments to the microscopic features in the optically recorded microstructure. We use the Mori-Tanaka micromechanical model \citep{mori1973average} to predict the effective bulk and shear moduli of the composite-like ``matrix" (plagioclase, olivine, and pyroxene) phase.

\begin{figure}[h!]
	\centering
	\includegraphics[width=0.5\textwidth]{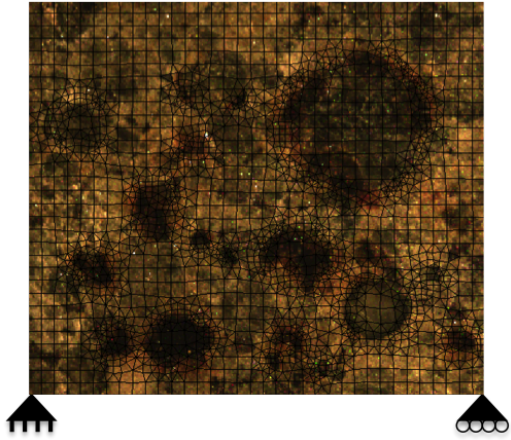}
	\caption{ Finite-element mesh generated from microstructure, together with the boundary conditions applied for the thermomechanical analysis.}
	\label{fig:oof_mesh}
\end{figure}

The effective thermal expansion coefficients (CTE) are calculated following Levin's \citep{levin1968estimating} relationship between the effective expansion coefficient and the effective elastic moduli of two phases, which were re-derived to incorporate the effective bulk modulus as well \citep{rosen1970effective}. Then, appropriate boundary conditions were applied. We use a Dirichlet boundary condition of traction-free boundaries, and we fix the longitudinal and transverse displacements at the bottom-left corner node to avoid numerical convergence problems due to unrestricted rigid body motion. No external forces were imposed on the system, and an initial nodal temperature corresponding to the room temperature was assigned. Figure~\ref{fig:oof_mesh} shows the resulting generated mesh with the applied boundary conditions.

It is worth noting that in the current configuration, the sample is allowed to expand freely so that no global stresses would develop from over-constraining the body's expansion. The thermal stresses that develop within the sample are solely due to the mismatch in mechanical and thermophysical properties among the constitutive phases, which lead to natural internal constraints on the deformation of each phase. A perfect bonding between phases and matrix was assumed. \\

In these calculations, the overall temperature is  gradually ramped up to 200\degree C, and the nodal displacement solutions are obtained for different temperature increments. Strains are then obtained from the displacements, and the computed strain field is compared to the experimentally measured strain fields to see if a match exists between the simulation results  and the experimental measurements. To evaluate the microscopic thermoelastic stresses that arise over the course of a solar day within each phase of the microstructure, the thermomechanical model is used to solve the equilibrium equation for the sample with traction free boundary conditions. The effect of temperature in this thermoelastic problem is incorporated into the stress tensor as:

\begin{equation}
\sigma_{ij}=C_{ijkl}(\epsilon_{kl}-\alpha \delta_{kl}\Delta T)
\end{equation}

\noindent where $\bold{C}$ is the stiffness tensor, $\epsilon$ is the strain tensor, $\alpha$ is the coefficient of thermal expansion in an element, $\delta_{kl}$ are the components of the identity tensor and $\Delta T$ is the increment in temperature from the reference room temperature. The solution converged with a relative residual error of $1\times10^{-14}$ using the conjugate gradient method. \\

\section{Computational results and comparison with experiment}
\label{sec:modelresults}
\begin{figure}[h!]
	\centering
	\includegraphics[width=1.0\textwidth]{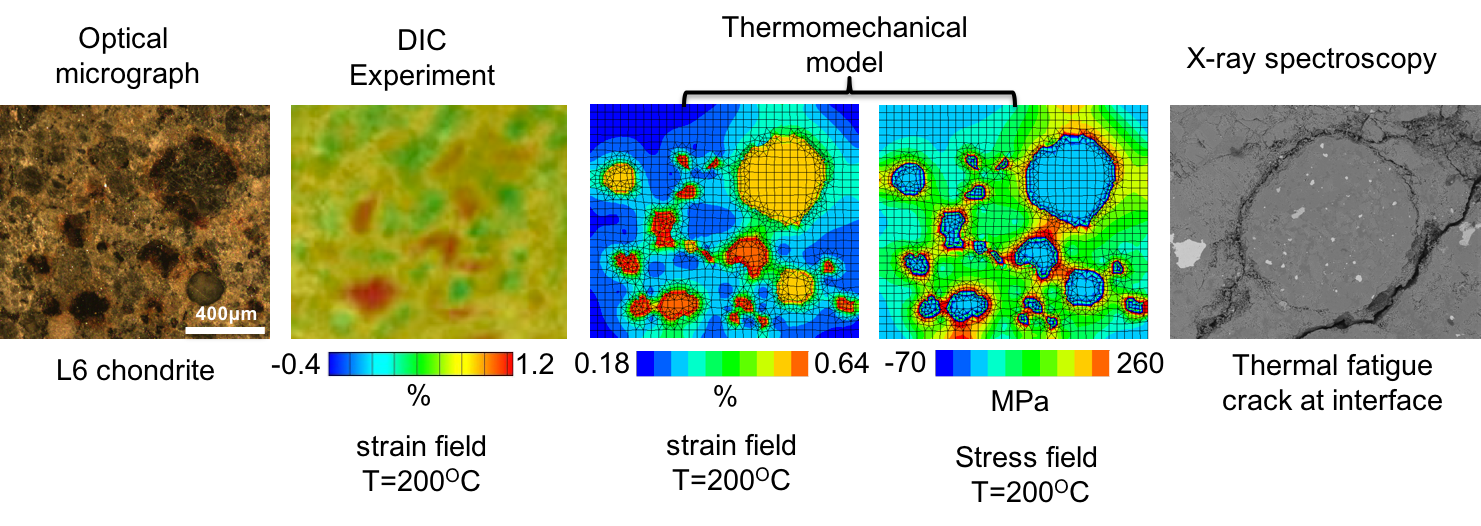}
	\caption{Comparison of experiment and simulation. The figure on the extreme left is an optical micrograph of a region of the sample. The next figure shows the experimentally measured strain distribution in that microstructure, using DIC. The middle figure shows the computed strains using our microstructurally-informed numerical model, showing good correspondence to those that were measured using the DIC technique.  The next figure shows the computed stresses. While the strains have an extrema within the inclusions, the stresses concentrate at the inclusion/matrix interface. These stress concentrations constitute favorable pathways for crack growth. Finally, an optical micrograph of the crack growth through the interfaces provide further evidence in support of this conclusion. Illustrating the meteorite microstructure and its associated major principal strain and stress obtained from the thermomechanical model ({\it cf. main text).}  Material and thermal coupling parameters assigned to all constituents are given (Table 1).}
	\label{fig:stress}
\end{figure}

The modeling results are shown in Figure~\ref{fig:stress}. This figure shows the original microstructure, the associated mesh generated for the simulations, the experimental principal strains measured using the DIC, the computed principal strain field for comparison, and the computed major principal stress field. The excellent comparison between the measured strain field and the computed strain field gives us confidence in the simulation itself, and thus we have confidence in the computed stress field.

As previously noted, high thermal strains do not necessarily correlate to high thermal stresses. In our case, the thermal stresses arise from the mismatch in thermal expansion coefficients between the inclusions and the matrix. Indeed, the calculated stresses appeared to be concentrated at the inclusion/matrix interfaces. In particular, during a thermal cycle from $100^o$C to $200^o$C, we find that the stress can vary from $0$ to $100$ MPa in the bulk phases, and from $0$MPa to $260$MPa at typical interface locations. 

\section{Discussion}
\label{sec:discussion}

The high stress levels calculated at the interfaces are important from a fracture mechanics viewpoint as they determine the crack propagation path. Indeed, we find in the experiments that the crack grows along the interfaces between the inclusions and the matrix, precisely the regions of the highest thermal stresses arising from the heterogeneous microstructure. There are two consequences. First, the resulting fragments will tend to be rounded, because the associated phases and interfaces have a rounded character. Second, a substantial fraction of the fragments are likely to be monomineralic since it is the interfaces that determine the crack path. Further, it is important to recognize that the overall fracture resistance and the crack growth rate is also affected by such compositional and morphological heterogeneities, spatial distributions, and the corresponding fracture process zones. Here we have focused on the crack path rather than the crack growth rate, which requires a more sophisticated simulation of the growing crack problem, to be described in a forthcoming work. Another important observation is that the stresses are themselves very heterogeneously distributed across the microstructure (see Figure~\ref{fig:stress}). This is perhaps correlated to the variability of times to fracture~\cite{delbo2014thermal} for typical heterogeneous materials such as L-chondrites. 

\begin{figure}[h!]
	\centering
	\includegraphics[width=1\textwidth]{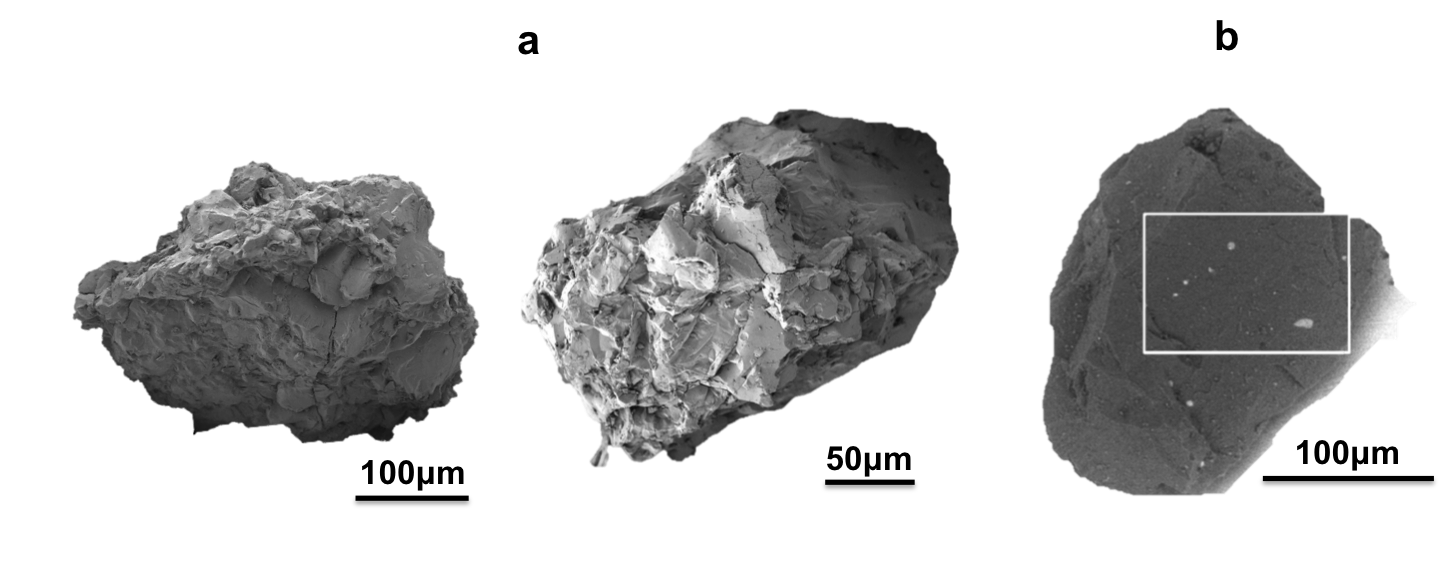}
	\caption{ Comparison between a) lab generated fragments and b) fragment collected by the JAXA Hayabusa mission  \citep{tsuchiyama2011three}. Impact generated fragments show distinct multi-facetad cleavage surfaces that are different from the smooth surface features of the fragments collected by Hayabusa. The thermal fatigue process we describe would generate fracture surfaces similar to the mono-mineral fragments collected from the surface of Itokawa}
	\label{fig:shapes}
\end{figure}

One of the significant consequences of the observed crack path in this thermally cycled chondrite is that it is likely that the importance and efficiency of the thermal fatigue mechanism is a strong function of the microstructure of the specific material considered (at least at these cm length scales - at larger length scales, as Delbo et al. pointed out, it is the thermal gradient that counts rather than the local microstructural heterogeneity). Our results suggest that boulders with few chondrules and iron-nickel phases may be less prone to disaggregation due to thermal fatigue. Further, our experimental evidence (Figure~\ref{fig:strainmaps}) together with the stress analyses obtained through our data-driven thermomechanical model (Figure~\ref{fig:stress}) suggests that thermal fatigue processes promote intergranular fracture. This is supported by the observation of the large number of mono-minerals \citep{nakamura2011itokawa} in the Itokawa fragments as reported from samples returned by the Hayabusa mission. Our results can thus perhaps be applied to differentiate  ({\it cf.} Figure~\ref{fig:shapes}) fragment generation resulting from thermal fatigue as opposed to fragmentation as a consequence of high velocity impact onto the surface of asteroids.  Impact generated fragments show distinct multi-faceted cleavage surfaces that are different from the smooth surface features of the fragments collected by Hayabusa. Note that Tsuchiyama et al. \citep{tsuchiyama2011three} suggested abrasion due to seismic-induced grain motion as a possible mechanism for the observed features. We note that there is nothing in our work that would suggest that this does not remain a feasible mechanism. However, the thermal fatigue process that we have demonstrated would not only generate fracture surfaces similar to the fragments collected from the surface of Itokawa, but would also result in the mono-mineral fragments that are observed.

Finally, we note that this crack propagation mechanism is also a possible mechanism for modifying existing regolith size and shape distribution on such bodies, leading to surface rejuvenation, both in terms of overall particle distributions on asteroidal surfaces and in terms of the development of relatively young surfaces on larger rocks.

\section{Summary}
\label{sec:summary}

In summary, we have performed a detailed mechanical characterization of an L6 ordinary chondrite specimen to obtain information concerning its constituents and their corresponding thermomechanical properties. The sample contained a pre-existing natural crack, which provided us with an opportunity to observe the crack growth path during several hundred thermal cycles. The sample was subjected to heating/cooling cycles, while simultaneously measuring the full-field strain map using a DIC setup. Using high-resolution images of the microstructure, we generated a representative mesh of the specimen and assigned the experimentally-measured material properties to the corresponding mesh elements. The thermally-induced stresses were then calculated, and the results indicated high stress concentrations at the inclusion interfaces. The observed stress distribution is indicative of the role that inclusions could play in dictating the fatigue crack growth path and the nucleation sites for new fatigue cracks. \\

Our experimental and theoretical investigations show that cracks can grow in chondrites through thermal fatigue, and that the crack path is primarily along interfaces between phases. This is potentially a principal source of fragments and particles on small airless bodies. This is also a possible mechanism for modifying existing regolith size and shape distribution on such bodies, leading to surface rejuvenation. Because of the strong effect of microstructure, it is likely that this mechanism's importance and efficiency is different for different materials, particularly at small (cm) scales.

Finally, we believe that crack growth due to thermal fatigue could, in principle, generate monomineralic fragments -- such as those recovered from Itokawa -- as a result of  crack growth around the mineral interfaces.

\section{Acknowledgements}
The authors would like to thank Solar System Exploration Research Virtual Institute-NASA for providing financial support for the current study. This study could not have been done without Smithsonian National Museum of Natural History support by providing meteorite samples. We would like to extend our special thanks to J. Hoskin, Collection Manager, and Dr. T. McCoy, Curator-in-Charge, Smithsonian Department of Mineral Sciences; Division of Meteorites. The Authors would like to thank Theoretical and Applied Mechanics Group at Drexel University led by Prof. A. Kontsos for assisting with measuring the strain maps. The authors would also like to thank Dr. S. Langer and Dr. A. Reid at the Center for Theoretical and Computational Materials Science at National Institute of Standards and Technology (NIST) for assisting with developing the thermomechanical model through the use of OOF2. Finally, we would like to thank Dr. O. Barnouin, Dr. J. Plescia, and Dr. A. Stickle at Applied Physics Laboratory for constructive discussions and feedback during this investigation. The work of M. Delbo was supported by the French Agence National de la Recherche (ANR) SHOCKS.


\end{document}